\documentclass{appolb}
\usepackage{graphicx}
\usepackage{braket}

\begin{document}
\title{Nuclear structure calculations in $^{20}$Ne with No-Core Configuration-Interaction model %
\thanks{Presented at the Zakopane Conference on Nuclear Physics {\em Extremes of the Nuclear
Landscape}, Zakopane, Poland, August 28 - September 4, 2016}%
}
\author{Maciej Konieczka and Wojciech Satu\l a
\address{Faculty of Physics, University of Warsaw, PL 02-093 Warszawa, Poland}
}
\maketitle
\begin{abstract}

Negative parity states in $^{20}$Ne and Gamow-Teller strength distribution for the ground-state beta-decay of $^{20}$Na 
are calculated for the very first time using recently developed No-Core Configuration-Interaction model. The approach is based on 
multi-reference density functional theory involving isospin and angular-momentum projections. Advantages and shortcomings 
of the method are briefly discussed.

\end{abstract}
  
\section{Introduction}
Single-reference  nuclear Density Functional Theory (SR-DFT) is a tool of choice  for large-scale calculation of bulk nuclear properties like masses, radii or quadrupole moments over the entire nuclear chart, see for example Ref.~\cite{[Erl12]} and refs. quoted therein. Despite many incontestable successes, the SR-DFT method has serious drawbacks.   It describes nuclei in the intrinsic frame of reference what inevitably leads to spontaneous breaking of fundamental symmetries in the mean-field wave function (Slater determinant). Hence, the method cannot be directly applied to account for structure and transitions in a
quantum-mechanically rigorous way. Increasing a range of applicability of the nuclear DFT requires, therefore, a restoration of relevant symmetries what can be done by applying projection 
techniques. It leads, after utilizing the generalized Wick's theorem, to a multi-reference DFT (MR-DFT) having 
the same form of the energy density functional (EDF) like the underlying SR theory but expressed in terms of transition densities.    
If the SR-EDF is generated by one of the most popular density-dependent Skyrme or Gogny forces, the 
projection leads, however, to singularities in the energy kernels and the entire framework becomes ill defined.  Attempts to overcome this difficulty by either directly 
regularizing the kernels~\cite{[Ben09],[Sat14b]} or constructing finite-range regularizable functionals \cite{[Rai14]} did not bring satisfactory solutions so far. 
At the moment, the only numerically stable realization of the MR-DFT scheme  is possible with the density-independent Skyrme SV ~\cite{[Bei75]} or 
SLyMR0~\cite{[Sad13b]} forces which are {\it true\/} interactions. These forces are characterized by anomalously low effective mass which is necessary to 
obtain correct saturation property. This, in turn, affects single-particle (SP) level density which scales linearly with the inverse of isoscalar effective mass.  
Whether or not this specific property will affect the spectra and transitions calculated using MR-DFT-rooted approaches is not obvious and is one 
of the goals of the present study.

The MR-DFT theories can be extended towards no-core configuration-interaction (NCCI) approaches by mixing states projected from the Slater 
determinants representing (many)particle-(many)hole  (or quasiparticle) configurations relevant for a given physical problem. 
The DFT-rooted NCCI  method, hereafter called NCCI for a sake of simplicity, was recently used to compute rotational spectra and electromagnetic transitions in $^{25}$Mg~\cite{[Bal14b]} and 
beta-decay matrix elements~\cite{[Sat14],[Kon16]}. 
The first applications are extremely promising and highly encouraging. Naturally, more 
systematic studies of NCCI models are needed before trumpeting full success. The aim of this work is to present further tests of the NCCI approach developed recently by our group, see Ref.~\cite{[Kon16],[Sat15],[Sat16]}. The model 
will be applied to calculate negative parity states in $^{20}$Ne and the Gamow-Teller strength for the ground-state (GS) beta-decay of $^{20}$Na.

\section{No-Core Configuration-Interaction Model}

The MR-DFT approach developed by our group uses a unique combination of the isospin and angular-momentum projections ~\cite{[Sat12]}. It provides wave functions projected from 
arbitrary (many)particle-(many)hole self-consistent mean-field configuration $\ket{\varphi}$:
\begin{equation}
\ket{\varphi;\, IM;T_z}^{(j)} = \frac{1}{\sqrt{\mathcal{N}^{(j)}_{IM;T_z}}}\sum_K \sum_{T\geq T_z}
a_{KT}^{(j)} \hat{P}^{T}_{T_zT_z}\hat{P}^{I}_{MK}\ket{\varphi},
\end{equation}
where the $\hat P^I, \hat P^T$ operators stand for projection operators of a SU(2) group generated by angular-momentum and isospin, respectively. 
The wave functions are labelled by quantum numbers of the angular-momentum, its projection onto $z$-axis, and the projection of isospin onto the $z$-axis in isospace. In this work we consider the Coulomb interaction (including exchange term), therefore the isospin $T$ is not a conserved symmetry of the Hamiltonian. 


The configuration space of the NCCI model used hereafter is built in the following way. First, a set of $n$ configurations (Slater determinants) $\ket{\varphi_i},\, i=1, 2, \ldots , n$  is calculated 
self-consistently using the Hartree-Fock (HF) method with the $\textrm{SV}_{\textrm{so}}$ force which is a variant of the SV parametrization with the strength of the spin-orbit term 
increased by 20\%, see Ref.~\cite{[Kon16]}.  In the set we include the GS and low-lying particle-hole (p-h) excitations of a given parity and signature as these 
symmetries are conserved at the HF stage of our calculations. Thus, both the aligned ($|h\rangle\otimes |\tilde p\rangle$ or $|\tilde h\rangle\otimes |p\rangle$ ) and antialigned 
($|h\rangle\otimes |p\rangle$ or $|\tilde h\rangle\otimes |\tilde p\rangle$)  p-h excitations are taken into account. Moreover, in the $N=Z$ nuclei, due to the symmetrization caused by isospin projection, we limit ourselves to 
neutron p-h configurations. It means that, in  the case of  $^{20}$Ne, the total number of linearly independent positive parity  p-h configurations covering the $sd$ shell is equal ten.
Let us underline that our p-h configurations involve excitation among deformed Nilsson states, see Fig.~\ref{Fig:F1}, which are mixtures of many
(multi)particle-(multi)hole spherical shell-model (SM) configurations. 

Eventually, the states  $\ket{\varphi_i ;\, IM;T_z}$ are mixed. At this stage we solve the Hill-Wheeler equation using the same $\textrm{SV}_{\textrm{so}}$ Skyrme interaction 
(plus Coulomb) in order to obtain the eigenenergies and the associated eigenfunctions: 
\begin{equation}
E^{(k)}_ {IM;T_z} \quad \textrm{and} \quad  \ket{IM;T_z}^{(k)}=\frac{1}{\sqrt{\mathcal{N}_{IM;T_z}^{(k)}}}\sum_{ij} {a_{ij}^{(k)} \ket{\varphi_i;IM;T_z}^{(j)}}.
\end{equation}
For further details we refer reader to Ref.~\cite{[Sat16]} .  


\section{Numerical results in $^{20}$Ne}

\subsection{Cross-shell excitations and negative parity states}

The DFT-rooted NCCI approach is, by construction, capable to describe both the positive and negative parity states using single universal two-body interaction.  It takes into account 
core-polarization or deformation effects inheriting them from the underlying mean-field treatment. This is of paramount importance for investigating 
cross-shell excitations to the intruder
or high-$j$ orbits which polarize or often statically deform a nucleus.

\begin{figure}[htb]
\centerline{
\includegraphics[width=10cm]{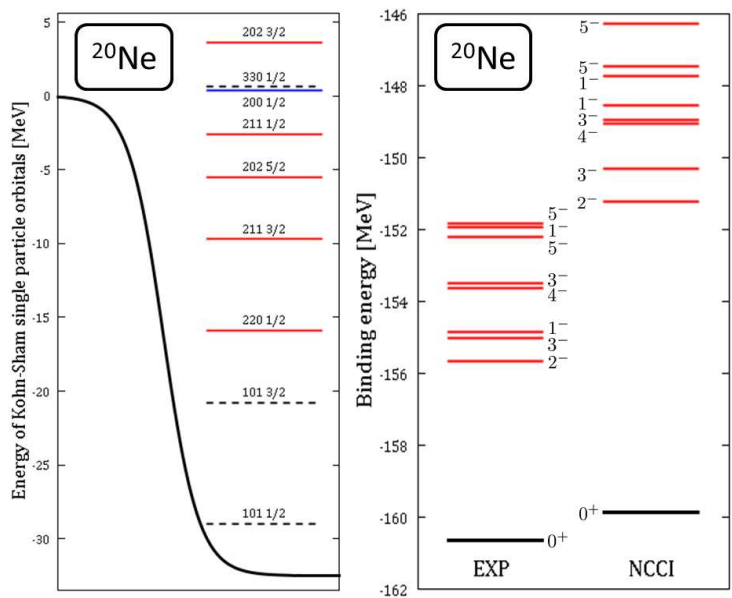}}
\caption{(Color online). Left part shows neutron SP levels in the GS of $^{20}$Ne. The orbitals are labelled with approximate Nilsson quantum numbers. Right part shows
the calculated spectrum of negative-parity states in $^{20}$Ne in absolute energy scale. See text for further details.}
\label{Fig:F1}
\end{figure}

An example of such calculations involving cross-shell p-h excitations in $^{20}$Ne is shown in Fig.~\ref{Fig:F1}.  In this preliminary calculations we limited the NCCI configuration space 
to two negative parity p-h configurations (see left panel Fig.~\ref{Fig:F1}), which correspond to aligned and antialigned excitations from the negative parity Nilsson level  [101]3/2 to the lowest available 
positive parity Nilsson level [211]3/2, see the left part of Fig.~\ref{Fig:F1}.  In the positive parity subspace we have mixed states projected from 
the GS and 10 p-h excitations
discussed above. Comparison between the calculated negative parity states and the data shows that we  
systematically overestimate the excitation energy of the negative-parity states by $\sim 3$\,MeV.  The primary reason can be traced back to a large,
$\sim 10$\,MeV,  energy splitting between the [101]3/2 and [211]3/2 Nilsson levels in the GS configuration. 
The polarization effect (or an increase of deformation) that accompany the p-h excitation reduces this energy by $\sim 2$MeV  but the effect is too small to account for experimental data.  
Note however, that the order and the spacing of the calculated negative parity states is in surprisingly good agreement with the data except
for the $1^-$ states. It remains to be studied how this picture would evolve in function of a number of negative parity configurations . 

\subsection{Gamow-Teller strength distribution}

Recently, we performed the first systematic study of the GS Gamow-Teller (GT) matrix elements in $T=1/2$ mirror nuclei using DFT-rooted models~\cite{[Kon16]}. It was shown that the MR-DFT calculations involving only one configuration representing the GS provides very 
accurate description of the matrix elements and that the results are stable against mixing additional configurations. However, in order to account for transitions to the excited states, 
especially to the configurations involving particles occupying close-lying SP levels, the mixing through the NCCI is indispensable.

\begin{figure}[htb]
\centerline{
\includegraphics[width=10cm]{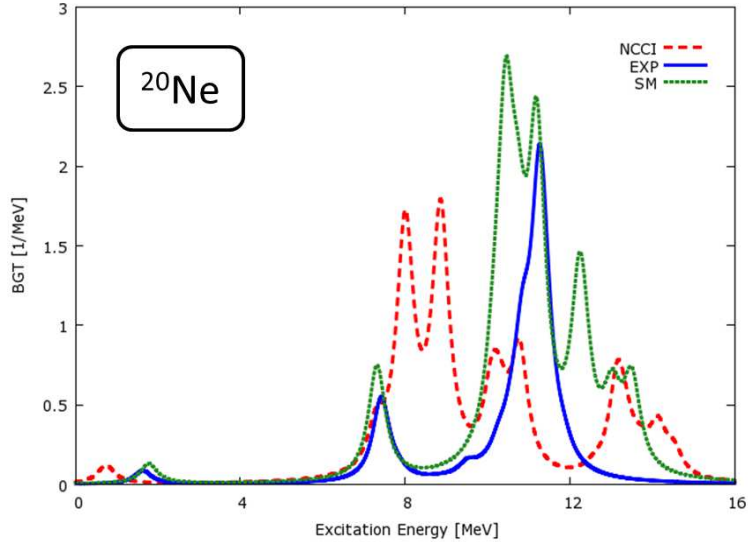}}
\caption{(Color online) Gamow-Teller strength function in $^{20}$Ne smeared by means of the Lorentz function of half-width 0.5\,MeV. Solid line labels experimental data. The SM calculations 
of Ref.~\cite{[Bro85]} are depicted by dotted line. Dashed line marks the DFT-rooted NCCI results.}
\label{Fig:F2}
\end{figure}

In this section we perform the calculation of  GT strength function for the GS decay of $^{20}$Na. The $I=2^+$ GS of  $^{20}$Na was calculated using the NCCI model involving four lowest-lying p-h configurations. The distribution of  $I=1^+, 2^+$ and 3$^+$ states  in the daughter nucleus $^{20}$Ne was computed  using the GS and 10 p-h configurations as discussed above.  
The GT strength function is presented in Fig.~\ref{Fig:F2} in comparison with the USDb SM results of Ref.~\cite{[Bro85]} and experimental data. The curves are smoothed using Lorentzian 
distribution with the half-width $\Gamma=0.5$MeV. The calculations show that the NCCI strength distribution is shifted towards lower energies with respect to both the SM calculations and experimental data. In particular, the resonant pick is shifted  by $\sim 3$MeV. This indicates that the GT strength distribution is very sensitive to 
the positions of specific SP Nilsson levels in a potential well. Thus, it may serve as an excellent tool to attempt to identify the correct order and spacing of orbitals in a given nucleus. It seems that the SP structure and in turn the GT strength distribution can be improved by increasing the 
spin-orbit strength. Local optimization of the interaction is, however, beyond the scope of this work.

\section{Summary and outlook} 
We presented very preliminary results of the DFT-rooted NCCI calculations for negative-parity states in $^{20}$Ne and the GT strength for GS decay of $^{20}$Na. It is shown that 
general features of experimental data can be captured with relatively limited number of p-h configurations. We encountered problems in reproducing the excitation 
energy of the first  $2^-$ state or the position of resonant picks in the GT distribution. These quantities are very sensitive to the positions of SP levels
in the mean-field potential well. Preliminary results indicate that an increase in  the strength of spin-orbit interaction may considerably improve the results.

This work was supported by the Polish National Science Center (NCN) under Contract No. 2014/15/N/ST2/03454. We acknowledge the CIS-IT National Centre for Nuclear Research (NCBJ), Poland and CSC-IT Center for Science Ltd, Finland for allocation of computational resources.       

\bibliographystyle{unsrt}
\bibliography{20Ne,jacwit33}

\begin{thebibliography}{10}

\bibitem{[Erl12]}
{J. Erler, N. Birge, M. Kortelainen, W. Nazarewicz, E. Olsen, A. M. Perhac, and
  M. Stoitsov, Nature {\bf 486}, 509 (2012)}.

\bibitem{[Ben09]}
{M. Bender, T. Duguet, and D. Lacroix, Phys. Rev. C {\bf 79}, 044319 (2009)}.

\bibitem{[Sat14b]}
{W. Satu{\l}a and J. Dobaczewski, Phys. Rev. C {\bf 90}, 054303 (2014).}

\bibitem{[Rai14]}
{F. Raimondi, K. Bennaceur, and J. Dobaczewski, J. Phys. G: Nucl. Part. Phys.
  {\bf 41}, 055112 (2014)}.

\bibitem{[Bei75]}
{M. Beiner, H. Flocard, N. Van Giai, and P. Quentin, Nucl. Phys. {\bf A238}, 29
  (1975)}.

\bibitem{[Sad13b]}
{J. Sadoudi, T. Duguet, J. Meyer, and M. Bender, Phys. Rev. C \textbf{88},
  064326 (2013)}.

\bibitem{[Bal14b]}
{B. Bally, B. Avez, M. Bender, and P.-H. Heenen, Phys. Rev. Lett. {\bf 113},
  162501 (2014).}

\bibitem{[Sat14]}
{W. Satu{\l}a, J. Dobaczewski, M. Konieczka, and W. Nazarewicz, Acta Phys. Pol.
  {\bf B45}, 167 (2014)}.

\bibitem{[Kon16]}
{M. Konieczka, P. B{\c a}czyk, and W. Satu{\l}a, Phys. Rev. C {\bf 93},
  042501(R) (2016)}.

\bibitem{[Sat15]}
{W. Satu\l{}a, J. Dobaczewski, and M. Konieczka, JPS Conf. Proc. {\bf 6},
  020015 (2015).}

\bibitem{[Sat16]}
{W. Satu{\l}a, P. B{\c a}czyk, J. Dobaczewski, and M. Konieczka, Phys. Rev. C
  {\bf 94}, 024306 (2016)}.

\bibitem{[Sat12]}
{W. Satu{\l}a, J. Dobaczewski, W. Nazarewicz, and T.R. Werner, Phys. Rev. C
  {\bf 86}, 054316 (2012)}.

\bibitem{[Bro85]}
{B.A. Brown and B.H. Wildenthal, Atomic Data and Nuclear Data Tables {\bf 33},
  347-404 (1985).}

\end{thebibliography}

\end{document}